# A Search for Extended Radio Sources in 1.3 sr of the VLA Sky Survey (VLASS)

Agueda Constanza Villarreal Hernández[1] and Heinz Andernach[2]

## RESUMEN

Reportamos resultados preliminares de una inspección visual de 4414 imágenes, cubriendo ~4260 grados cuadrados del rastreo a 3 GHz VLA Sky Survey (VLASS, época 1.1) en búsqueda de estructuras extendidas. Se colectaron más de 7600 posiciones y para una submuestra de 270 de las candidatas más prometedoras se buscó el objeto huésped en rastreos ópticos e infrarrojos, tal como en otras bases de datos públicas. Entre las 270 encontramos 9 nuevas radiogalaxias gigantes con una extensión en radio mayor a 1 Mpc, dos de ellas probablemente cuásares a un corrimiento al rojo de z~2, y otra estuvo identificada hasta ahora erróneamente con un objeto más cercano. Aparte de varias fuentes con una radiomorfología inusual, encontramos numerosas fuentes del tipo remanente con lóbulos muy difusos en VLASS, tanto con o sin un radionúcleo. Confirmamos que la alta resolución angular y frecuencia de observación de VLASS son idóneos para la identificación óptica de radiofuentes, y a pesar de su limitada sensibilidad para estructuras extendidas, permite revelar radiogalaxias grandes basándose en sus rasgos morfológicos, irreconocibles en rastreos de peor resolución angular.

## ABSTRACT

We report preliminary results of a visual inspection of ~4300 deg$^2$ covered by 4414 images of the 3-GHz VLA Sky Survey (VLASS, epoch 1.1) in search of extended radio structures. Over 7600 positions were registered, and for a subset of 270 of the most promising candidates their host objects were searched in optical and infrared surveys, as well as in other public databases. Among these 270, we found 9 new giant radio galaxies with a projected radio extent exceeding 1 Mpc, of which two are likely quasars at a redshift of z~2, and another one had hitherto been misidentified with a low-redshift host. Apart from various sources with unusual radio morphologies, we found numerous remnant-type double radio sources with very diffuse lobes in VLASS, both with and without radio cores. We confirm that the high angular resolution and observing frequency of VLASS is well suited for the optical identification of radio sources, and despite its limited sensitivity to extended emission, allows to reveal large radio galaxies based on morphological features unrecognizable in lower-resolution surveys.

**Palabras Clave:** rastreos en radio, radiofuentes extragalácticas, radiogalaxias

## INTRODUCTION

The VLA Sky Survey (VLASS, https://science.nrao.edu/science/surveys/vlass, Lacy et al., in prep.) is being performed with the Jansky Very Large Array (VLA) in B-configuration at S band (2-4 GHz), giving an angular resolution of 2.5". From Sept. 2017 to Febr. 2018 (epoch 1.1) half of the sky north of DEC=−40° was observed, resulting in 17,538 publicly available quicklook (QL) images covering 16,831 deg$^2$. The 1-$\sigma$ noise of the QL images is ~0.13 mJy/beam, expected to reach 0.069 mJy/beam with three full sky coverages completed by 2024. While the QL images are preliminary products and will be replaced with the final "SingleEpoch" images, an inspection of ~2000 deg$^2$ of QL images around the North Celestial Pole by Andernach (2018) had shown that even the QL images have excellent capability to reveal morphological features of radio galaxies like radio cores, jets, hotspots with Mach-cone shaped tails, etc., which are hardly perceived in lower-resolution surveys like the NRAO VLA Sky Survey (NVSS, Condon et al. 1998), or the TGSS-ADR1 (Intema et al. 2017). Since recognition of such structures with automated algorithms is challenging, this has motivated us to inspect further large areas of VLASS. For the present project we selected 9 contiguous areas located at Galactic latitudes |b|>10°, except for some small regions near Galactic longitudes 58°, 194°, and 214°. These 9 areas also have as little overlap with the FIRST survey (Helfand et al. 2015), since the latter has been explored extensively for optical identification of radio galaxies, e.g. within the Radio Galaxy Zoo project (Banfield et al. 2015). Table 1 lists for each area its RA and DEC limits, the number of available VLASS images, followed by the number of images rejected for quality issues, the area covered by the available images, the number of positions logged and their average number per image. Note that parts of the quoted RA-DEC ranges will be observed in epoch 1.2 in 2019.

[1] Universidad de Guanajuato, Div. Ciencias Naturales y Exactas, Guanajuato, Gto, a_agueda@hotmail.com
[2] Universidad de Guanajuato, Depto. de Astronomía, DCNE, Guanajuato, Gto, heinz@astro.ugto.mx

*Table 1. Sky areas inspected in VLASS*

| Area | RArange/h | DECrange/° | $N_{img}$ | $N_{reject}$ | A/deg$^2$ | $N_{log}$ | $N_{log}/N_{img}$ |
|---|---|---|---|---|---|---|---|
| 1 | 00 .. 02 | -10 .. -24 | 379 | 1 | 363.7 | 593 | 1.6 |
| 2 | 02 .. 04 | -24 .. -36 | 334 | - | 311.2 | 408 | 1.2 |
| 3 | 00 .. 03 | +10 .. +16 | 268 | 2 | 261.0 | 487 | 1.8 |
| 4 | 03 .. 06 | +00 .. +16 | 719 | 1 | 709.7 | 962 | 1.3 |
| 5 | 03 .. 07 | -00 .. -24 | 1196 | 10 | 1162.4 | 2354 | 2.0 |
| 6 | 10 .. 12 | -10 .. -24 | 419 | 1 | 399.7 | 654 | 1.6 |
| 7 | 12 .. 14 | -24 .. -36 | 332 | 2 | 309.2 | 692 | 2.1 |
| 8 | 14 .. 16 | -10 .. -24 | 410 | 10 | 391.1 | 745 | 1.8 |
| 9 | 20 .. 24 | +10 .. +16 | 357 | 3 | 347.7 | 748 | 2.1 |
| Total/average | - | - | 4414 | - | 4255.7 | 7643 | 1.7 |

## METHOD

Each QL image covers an area of 62' x 62' with 3722$^2$ pixels of 1". To display the QL images we used the program ObitView (W. Cotton, http://www.cv.nrao.edu/~bcotton/Obit.html), which allows to record a position on an image with a single click, accumulating them in a file. For our search we used a 15.6" screen of 3480 x 2160 pixels at a zoom of 50%, which allowed us to view a large fraction of the full image per screen with sufficient detail. Given the limited time for this project, we deferred the measurement of angular sizes until later. Since extended radio sources are rare on VLASS images, selecting all except the more numerous double sources smaller than ~30" can be done within an average of 1 minute per image. ObitView permits two types of position logging: for compact radio cores we performed a Gauss-fit with sub-arcsecond precision on the source, while for candidates without an obvious core we clicked on a position nearest to the source's geometric center. For the most promising candidates (either large angular size or peculiar shape) we employed a double click, which could be distinguished easily from the other sources in the log file. After cleaning duplicates due to the small overlap of ~4' between neighbouring images, a total of ~7640 positions were logged on the 4414 QL images.

Of the 325 double-clicked candidates, 50 had already been compiled as extended radio galaxies by Andernach (2016). For the remaining 275 objects we inspected 15'x15' NVSS images to select 145 sources which appeared as genuine radio galaxies larger than ~1.5'. Using the program ds9 (http://ds9.si.edu) we measured their largest angular size (LAS), along a straight line between opposite extremes of their radio extent, on the VLASS images (or occasionally on NVSS or TGSS images, when they were almost resolved out in VLASS). We then searched for counterparts of their radio cores within ~2" in optical surveys (SDSS, Abolfathi et al. 2018; Pan-STARRS DR1, Flewelling et al. 2016) as well as in the mid-infrared AllWISE catalog (Cutri et al. 2013) using both the VizieR catalog browser (http://vizier.u-strasbg.fr/viz-bin/VizieR, Ochsenbein et al. 2000) and the NASA/IPAC Extragalactic Database (NED, http://ned.ipac.caltech.edu), resulting in matches for 105 of the 145 objects. Eight objects had a spectroscopic redshift ($z_{spec}$) and for the remaining ones we searched photometric redshifts ($z_{phot}$) for galaxies in catalogs by Beck et al. (2016), Brescia et al. (2014), Bilicki et al. (2016) and Gao et al. (2018). For star-like objects which, when radio-emitting, are highly likely to be quasars, we used redshift estimates from DiPompeo et al. (2015) and Krogager et al. (2018, their Fig. 2). We used $z_{spec}$, or, in its absence, an average of the $z_{phot}$ values, to convert the LAS to the largest linear size (LLS), projected on the sky, in Mpc (1 Mpc = 3.09 x 10$^{22}$ m =3.26 x 10$^6$ light years), adopting a standard cosmology with H$_0$=70 km s$^{-1}$ Mpc$^{-1}$, $\Omega_m$ =0.3 and $\Omega_\Lambda$=0.7.

## RESULTS AND DISCUSSION

Here we give only a few examples of extraordinary sources found. All images have N at the top and E on the left. Note that all sources were first seen on the VLASS images, and confirmed only *a posteriori* in NVSS or TGSS.

**Giant Radio Galaxies:** We found 11 objects to be giant radio galaxies (GRGs) having an LLS > 1 Mpc, of which only two were previously known. Another 20 sources with LLS > 0.7 Mpc were found. Figure 1 shows two of the largest GRGs. The source PKS 1449−129 (Fig. 1a) had previously been identified with 2MASX J14523554−1311144, rK=17.0m, at z=0.069. However, the radio core in the VLASS image is ~7.7" SW of the 2MASX galaxy (marked by a red ellipse) and coincides with the QSO candidate PSO J145235.301−131120.98, rK=19.59m, aka AllWISE J145235.30−131120.4, the bluish star-like object in the Pan-STARRS inset of 25" on a side. Its WISE colors suggest $z_{phot}$~2.2 (Krogager et al. 2018), thus its LAS~3.37' implies an LLS of ~1.7 Mpc. Figure 1b shows the VLASS image of GRG J0452+0247 with NVSS contours superposed. It is hosted by PSO

J045205.47+024748.8, the central one of a triplet of faint galaxies. No $z_{phot}$ is available, but with a host as faint as r=21.36m it should have at least z~0.5, thus its LAS=3.54' yields an LLS of at least 1.3 Mpc.

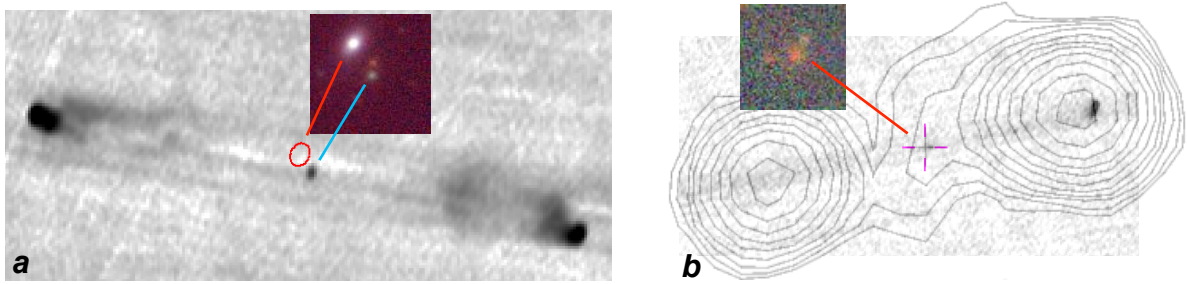

Figure 1. VLASS images of (a) J1452−1311 and (b) J0452+0247, with Pan-STARRS insets of 25" and 15"

Other examples of giants are two QSO candidates: (a) AllWISE J034739.23−271223.5, barely detected but uncatalogued in Pan-STARRS with IR colors suggesting $z_{phot}$~2.2, has LAS=2.2' and LLS=1.1 Mpc, and (b) PSO J204726.628+110418.93, rK=21.45m, with IR colors placing it at $z_{phot}$~0.7, has LAS=2.43', thus LLS=1.04 Mpc.

**Remnant double radio galaxies:** These have been evasive in past radio surveys requiring novel instruments like LOFAR to find them (Mahatma et al. 2018). Surprisingly, and despite the supposedly low sensitivity of VLASS to diffuse emission we found numerous pairs of faint and diffuse emission regions (lobes), sometimes with a central radio core indicating the location of the host. With very few exceptions, these are fairly nearby (z < 0.2), with an LAS < 3', and LLS < 1 Mpc. Two examples are shown in Fig. 2: (a) 2MASX J20305835+1013199, LAS=2.97', $z_{phot}$=0.02, LLS~73 kpc, and (b) 2MASX J22061299+1532107, LAS=2.5', $z_{phot}$=0.07, LLS~200 kpc. While only the latter one, lacking a radio core, complies with Mahatma et al's definition of a remnant, the absence of hotspots and presence of an active nucleus in the former may be due to recurrent (restarting) activity.

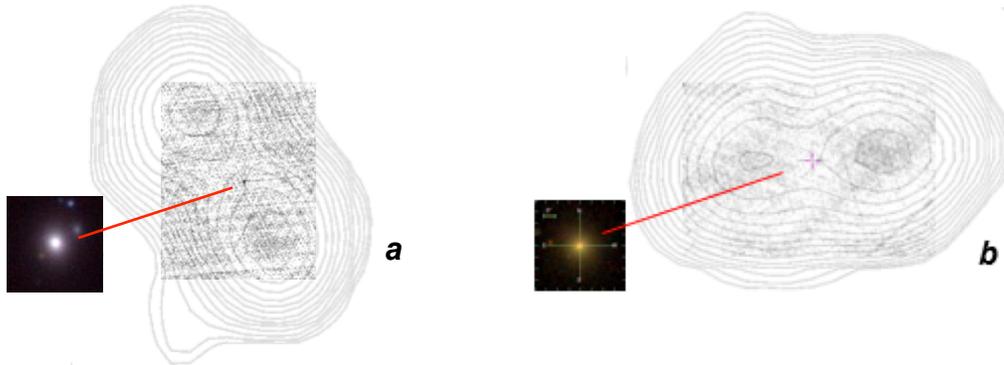

Figure 2. VLASS images of remnant radio galaxies (a) J2030+1013 and (b) J2206+1532, both with contours from NVSS, and insets of 20" from Pan-STARRS (a) and 40" from SDSS (b)

**Sources with peculiar shapes:** Figure 3a shows the VLASS image of J0342−0435 (4C−04.12, PKS 0339−047). Its SW lobe shows a jet-like feature, misaligned with the source major axis, terminating in a hotspot, and further bending backwards. Despite that it resides in a rather dense environment of a yet unreported X-ray emitting cluster (2RXS J034208.1−043539), it has LAS=3.28', $z_{phot}$=0.3 and thus LLS=0.88 Mpc. The rK=19.65m host galaxy PSO J034208.50−043542.1 is at the center of the 1' x 1' Pan-STARRS inset. Figure 3b shows the source J0105−2146 of LAS=0.98': a central core plus three hotspots, two bright ones to the north (H1, H2) and a weak one to the south. There is no optical or IR counterpart for any of its three hotspots. The core coincides with PSO J010545.63−214657.6 (rK=18.31m) aka AllWISE J010545.63−214657.5 with IR colors suggesting a quasar at z~1.5 (Krogager et al. 2018), implying LLS=490 kpc. There is no simple explanation for the N lobe: either the N jet, upon hitting the intergalactic medium in between hotspots H1 and H2, has deviated due both E and W, or the jet first created hotspot H1, followed by a ~90° bend due E, and then created hotspot H2, suggesting a total bend of ~180°. Figure 3c shows the E-W remnant-type source hosted by 2MASX J01062404−1326563 (LAS 2.5' from the TGSS-ADR1 image, $z_{phot}$=0.07, LLS=120 kpc) with a ring-shaped E lobe, while the W lobe is almost completely resolved out in VLASS. The 15" inset from Pan-STARRS shows the rK=15.17m elliptical host.

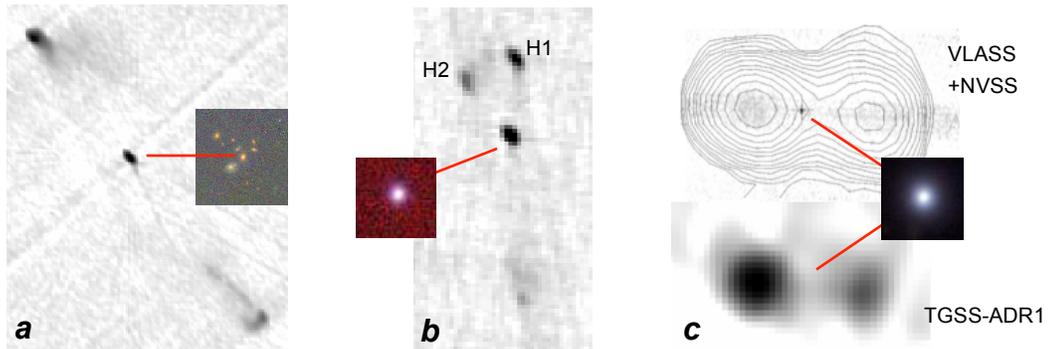

*Figure 3. VLASS images of (a) J0342−0435, (b) J0105−2146, and (c) J0106−1326 with contours from NVSS and a greyscale image of TGSS-ADR, and insets from Pan-STARRS of (a) 60″, (b) 10″ and (c) 15″ size*

The 1.6-GHz VLA image of the source PKS 1358−113, hosted by the brightest galaxy (MCG −02-36-002, $z_{spec}$=0.037) in the cluster Abell 1836 (Stawarz et al. 2014) shows a peculiar emission ridge at the SW border of its SE lobe (Fig. 4a). The VLASS image (Fig. 4b) suggests straight jets on opposite sides of the host. It also shows that the radio peak at the SSE end of the bright ridge in the SE lobe (Fig. 4c) coincides, as shown in the 12.5" inset from Pan-STARRS, with the r'=21.85m galaxy SDSS J140149.36−113904.2 which has a $z_{phot}$ of 0.37. This suggests that the peculiar emission ridge is actually due to a tailed radio galaxy (LAS~1.1', LLS~340 kpc) about ten times more distant than Abell 1836. Stawarz et al. (2014) classified PKS 1358−113 as ~6 times underluminous for its FRII morphology. Correcting now for the background source, PKS 1358−113 would be even more underluminous and asymmetric in lobe flux ratio, but more symmetric about its host.

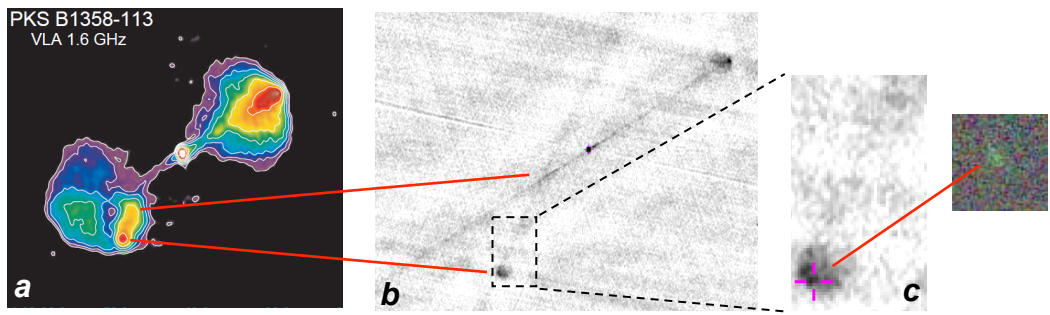

*Figure 4. PKS 1358-113 in Abell 1836 (a) at 1.6 GHz with VLA; (b) in VLASS; (c) SE ridge (see text)*

**The nature of the remaining candidates:** Finding optical counterparts for the remaining ~7300 candidates is well beyond the present project. A semi-automatic exploration gave the following results. For the ~3900 Gauss-fit positions (likely radio cores) we found ~100 matches within 1.5" of radio galaxy hosts in HA's compilation (15 of these GRGs). Of the remaining 3800, 1890 had matches in Pan-STARRS, and 490 in SDSS. Of these 490, 93 have a $z_{spec}$, 360 have $z_{phot}$ from Beck et al. (2016), 322 from Brescia et al. (2014), and 136 have a $z_{phot}$ among the QSO candidates in DiPompeo et al. (2015). Another 132 non-SDSS values of $z_{spec}$ were found in NED. Of the 3900 Gauss-fit positions, over 2300 have a match in AllWISE within 2.5", and of the latter, 664 have a $z_{phot}$ in Bilicki et al. (2016), and ~600 have IR colors ($W_{12} > 0.9$) indicative of quasars. Of the ~3400 non-Gauss-fit positions (sources without radio core in VLASS) only ~50 had a counterpart within 1' in the RG compilation of one of us (HA). We conclude that the vast majority of the remaining candidates, if real, will be new findings.

## CONCLUSIONS

Our 7-week project has shown that VLASS has a high potential of revealing large and unusual radio galaxies, as well as many others that would appear as inconspicuous and barely resolved double sources in low-resolution surveys, and that visual inspection is an efficient way to find these. We have registered the positions of extended radio sources in one quarter of all epoch-1.1 VLASS images within 75 net hours of work, implying that this task can be accomplished for the full 33,880 deg$^2$ of sky to be covered by VLASS in about 15 person-weeks. For many well-known powerful radio galaxies VLASS may reveal revised optical identifications, while many radio galaxies with undetected radio cores in VLASS can qualify as possible new remnant radio galaxies. To facilitate this type of research it is highly desirable that a postage stamp server for VLASS cutouts be offered, and that photometric redshifts for objects detected in Pan-STARRS were derived, as it covers more than twice the sky area than SDSS does. Our list of candidate positions may be useful for a variety of research projects, or as a training set for machine learning algorithms. Researchers interested in this list should contact the second author.

## ACKNOWLEDGEMENTS

We thank Amy Kimball (NRAO) for competent advice on technical details of the VLASS survey. We also thank her and Roger Coziol for useful comments on the manuscript, as well as Isabel Valdés Ochoa for help with the layout of figures. H.A. benefited from Universidad de Guanajuato grant DAIP #66/2018.